\documentclass[%
 reprint,
 amsmath,amssymb,
 aps,
 prd,
]{revtex4-2}

\usepackage{graphicx}
\usepackage{dcolumn}
\usepackage{bm}

\begin{document}

\preprint{APS/123-QED}

\title{Geodesic completeness from string T-duality}

\author{Kimet Jusufi$^{a}$}

\author{Piero Nicolini$^{b,c}$}%
\affiliation{%
$^a$Physics Department, University of Tetova, Ilinden Street nn, 1200, Tetova, North Macedonia 
\\
$^b$Dipartimento di Fisica, Università degli Studi di Trieste, and
Istituto Nazionale di Fisica Nucleare (INFN), Sezione di Trieste, Trieste, Italy\\
$^c$Institut für Theoretische Physik, Johann Wolfgang Goethe-Universität, and
Frankfurt Institute for Advanced Studies (FIAS), Frankfurt am Main, Germany}%

\email{piero.nicolini@units.it}


\date{\today}

\begin{abstract}
 By studying the Raychaudhuri equation for the gravitational force resulting from a string T-duality modified propagator, we present an analysis of the geodesic compression beyond the conventional classical limit.
The result is that gravity on short length scales is subject to a screening effect similar to the Debye screening in electrostatics, which prevents the formation of curvature singularities.  
Using model-independent arguments, we conclude that the conventional attractive nature of gravity is only a low-energy effect.
\end{abstract}

\maketitle



\section{Introduction}

From the discovery of Cygnus X-1 in 1964 \cite{BBCF65} to the more recent detection of gravitational waves \cite{GW15} and imaging of event horizons \cite{EHT19}, astronomical observations have confirmed the existence of black holes with increasing precision.
Despite the success of general relativity in predicting these phenomena, since the 1960s many authors have tried to overcome the problem of the curvature singularity for black hole spacetimes by invoking a variety of short-scale non-classical effects \cite{Gli66,Sak66}.
On the other hand, around the same years, Penrose - first alone \cite{Pen65} and later in collaboration with Hawking \cite{HaP70} - proved singularity theorems that left little room for the existence of regular black hole solutions.

Over time, however, some cracks gradually formed in the absolute belief in singularity theorems. For example, Wald \cite{Wal84} and, more recently, Kerr \cite{Ker23} have pointed out that the incompleteness of the geodesics is not in itself a sufficient condition for the existence of a curvature singularity. Similarly, Birrell and Davis pointed out in 1982 that in the presence of particle creation, the conservation of quantum stress tensors requires the violation of energy conditions
 \cite{BiD84}.
We also note here that since the years when these theorems were proved, black hole physics has drastically changed both theoretically and experimentally. First, the violation of energy conditions is no longer a theoretical speculation, but an experimentally observed phenomenon \cite{Lam97}.
Second, research in quantum gravity has made great progress.  For example, as early as the 1980s, studies in string theory showed that black holes represent the phase that matter assumes at energies higher than the Planck scale \cite{ACV87,ACV88,ACV89,tHo90}. In modern terms, this black hole reinterpretation, also known as ``classicalization'' \cite{AAS02,DFG11,DGGK11,Car13} or ``self-completeness'' of gravity \cite{DvG10,AuS13,MuN13,CMN15}, manifests itself as an exchange of length scales $R\to 1/R$, similar to the string T-duality \cite{Nic22,Nic23}.
The above two points lead the present discussion to the third: black holes may not simply be classical solutions of general relativity, as the growing literature on regular black holes suggests \cite{Bam23}. 

In order to shed light on the nature of black holes and ultimately on gravity itself, in this paper we aim to scrutinize the Raychaudhuri equation with a black hole metric modified by string T-duality (for short, ``string T-duality metric'') \cite{NSW19}. The motivations for the proposed investigation can be grouped as follows:
\begin{enumerate}
\itemsep0em 
\item The Raychaudhuri equation is at the heart of the above singularity theorems \cite{Raychaudhuri}; it provides insight into the attractive nature of gravity. 
\item The string T-duality metric is a real quantum-gravity corrected metric \cite{Nic23}; this is in contrast to many regular black hole models, where the connection to fundamental principles is missing or not transparent \cite{Dym92,MbK05,BrF06,Hay06}.
\item The string T-duality metric has universal properties since it depends on only one parameter, the string tension; therefore the results of the proposed investigation will be general in the sense that they hold within the whole class of regular black holes.
\item The string T-duality metric describes a neutral, static spacetime; therefore it does not suffer from the Schwinger instability of those regular metrics derived by coupling some non-Maxwell electrodynamics theory to gravity \cite{Bar68,AyG99c}.
\end{enumerate}


\section{String T-duality metric}
\label{sec:tdualitybh}

The string T-duality metric describes the gravitational field generated by a mass $M$ at the origin of the coordinates, under conditions of spherical symmetry.
In this case, the generic line element is
\begin{equation}\label{eq:metric1}
ds^2=-f(r)dt^2+\frac{dr^2}{f(r)}+r^2 d\theta^2+r^2 \sin^2\theta d\phi^2,
\end{equation}
where $f(r)$ must be determined by Einstein's equations.

Following the original derivation \cite{NSW19}, one starts by determining the static potential due to the virtual particle exchange between the source $M$ and a generic probe mass.  In doing so, one implements short scale quantum modifications using the T-duality modified propagator proposed by Padmanabahn, Spallucci and collaborators \cite{Pad97,SSP03,SpF05,FSP06}. The resulting Newton-Poisson equation no longer describes a point-like object. 
In fact, the mass density $\rho$ is distributed over a distance $l_0$, the so-called ``zero point length'', with $l_0=2\pi\sqrt{\alpha^\prime}$  where $\alpha^\prime$ is the Regge slope.
More importantly, the mass density is equal to $-T_0^{\ 0}$, i.e. the $00$-component of the energy-momentum tensor, which goes into the r.h.s. of Einstein's equations.
So the solution is
\begin{eqnarray}\label{eq:metriccoeff}
    f(r)=1-\frac{2Mr^2}{(r^2+l_0^2)^{3/2}},
\end{eqnarray}
after setting $G=1$. 
In the limit $r\gg l_0$ the spacetime tends to the Schwarzschild geometry.  For $M>3 \sqrt{3}\,l_0/4\simeq 1.30 \ l_0$, the horizon equation admits two solutions, an inner (Cauchy) horizon $r_-$ and an outer event horizon $r_+$. This means that the global structure is drastically different from that in the Schwarzschild case: Surfaces $r=0$ become time-like. In the limit of large masses $M\gg l_0$ the Chauchy horizon goes to the origin, while the event horizon approaches the Schwarzschild value $2M$.
On the other hand, for $M<3 \sqrt{3}\,l_0/4$ there is no horizon. Finally, for $M=3 \sqrt{3}\,l_0/4$ the two horizons merge into a single degenerate horizon.

One of the most spectacular properties of the metric \eqref{eq:metriccoeff} is that it is formally the Bardeen spacetime after replacing the magnetic charge with the zero point length $l_0$. This means that the above line element enjoys the good properties of the Bardeen solution - namely the absence of curvature singularity - but it does not require the existence of magnetic monopoles. More importantly, it is not a transient state because, unlike the Bardeen metric, it does not suffer from Schwinger instability \cite{Gib75,Pag06,Nic18}. 

\begin{figure}[ht!]
		\centering
	\includegraphics[scale=0.65]{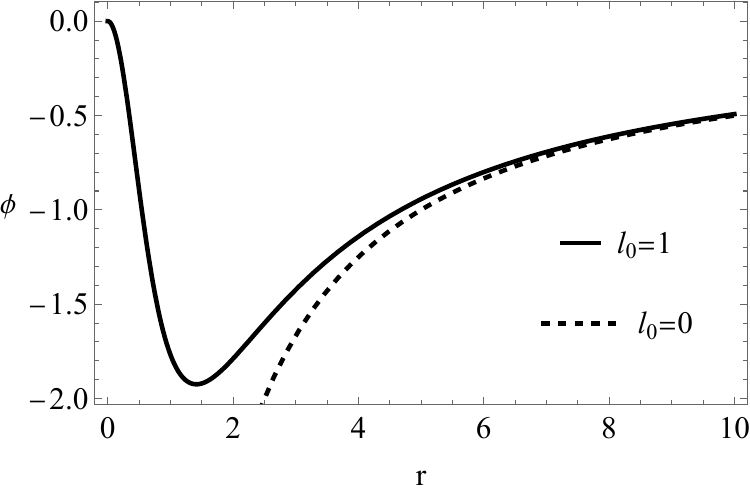}
\caption{Modified Newtonian potential $\phi(r)$ (solid line) versus standard Newtonian potential (dashed line) for a mass $M=5/{\sqrt{G}}$ with $G=1$.}
\label{fig:potential}
	\end{figure}

After these initial remarks, we introduce a modified Newtonian potential from the relation $f(r)=1+2\phi(r)$, with $\phi(r)=-m(r)/r$, where 
\begin{equation}
m(r)=\frac{Mr^3}{(r^2+l_0^2)^{3/2}}
\end{equation}
is the cumulative mass distribution. From Fig. \ref{fig:potential} we can see that $\phi(r)$ is attractive only at large distances, i.e. $r>1.7 \ l_0$. Conversely, in a region of size about $l_0$ around the origin, quantum fluctuations dominate and their mean value locally forms a repulsive de Sitter core.

\section{Geodesics in string T-duality metric}
\label{sec:geodcomple}


We start by recalling that for a generic observer $x^{\mu}_s(\tau)$ in Schwarzschild coordinates, the four-velocity $u_s^{\mu}\equiv dx^{\mu}_s(\tau)/d\tau$ is given by
\begin{eqnarray}
    u_s^{\mu}=\left(\frac{1}{f(r)},-\sqrt{1-f(r)}, 0, 0 \right).
\end{eqnarray}
Now consider a free-falling observer. For this purpose it is convenient to introduce the Painlev\'{e}--Gullstrand coordinates $(T, r, \theta, \phi)$, where the new time coordinate is \cite{Poisson:2009pwt}
\begin{equation}\label{Painleve time}
dT\equiv dt_s+\frac{\sqrt{1-f(r)}}{f(r)}dr\equiv d\tau .
\end{equation}
As a result, the metric \eqref{eq:metric1} becomes
\begin{equation}\label{painleve metric}
ds^2=-dT^2+\left(dr+v(r) \,dT \right)^2+r^2 d\Omega^2\,,
\end{equation}
where $v(r)=\sqrt{1-f(r)}$ and $T$ are the radial velocity and the proper time of a free-falling observer, respectively, assuming that the initial velocity vanishes at infinity.  We also note that there is no coordinate singularity at the horizon.

Consider the normalized vector field $u^a=(1, -v(r),0,0)$, namely the four-velocity in Painlev\'{e}--Gullstrand coordinates and a smooth congruence of time-like geodesics parametrized by the proper time $\tau$. Then the Raychaudhuri equation is
\begin{equation}
\frac{d\Theta}{d\tau} = - \frac{1}{3}\Theta^2 - \sigma_{\mu \nu } \sigma^{\mu \nu } + \omega_{\mu \nu }\omega^{\mu \nu }- R_{\mu \nu } u^{\mu} u^{\nu},~
\label{eq:rayeq} 
\end{equation}
where $\Theta$ is the volume expansion/contraction parameter defined as $\Theta=\nabla_{\mu}u^{\mu}$, $\sigma_{\mu \nu}$ is the shear tensor describing shape distortion without volume change, given by 
\begin{eqnarray}
    \sigma_{\mu \nu}=\left(\nabla_{\alpha} u_{(\mu}\right){P^{\alpha}}_{\nu)}-(\Theta/3)P_{\mu \nu} 
\end{eqnarray}
 where 
 \begin{eqnarray}
     P_{\mu \nu}=g_{\mu \nu}+ u_{\mu} u_{\nu},
 \end{eqnarray}
is the projector operator \cite{Carroll:2004st}.
Finally, the twist tensor $\omega_{\mu \nu}$ describes the rotation without volume change, but it vanishes for locally hypersurface orthogonal geodesics and therefore we will not consider it. 
 \begin{figure*}[ht!]
		\centering
	\includegraphics[scale=0.69]{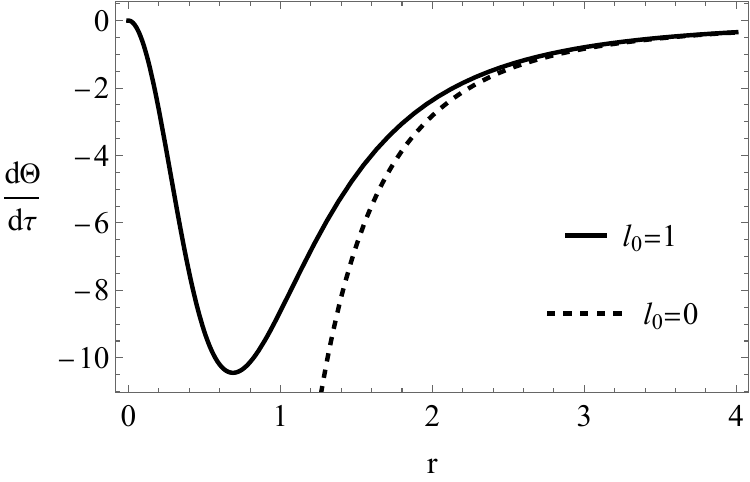}
 \includegraphics[scale=0.67]{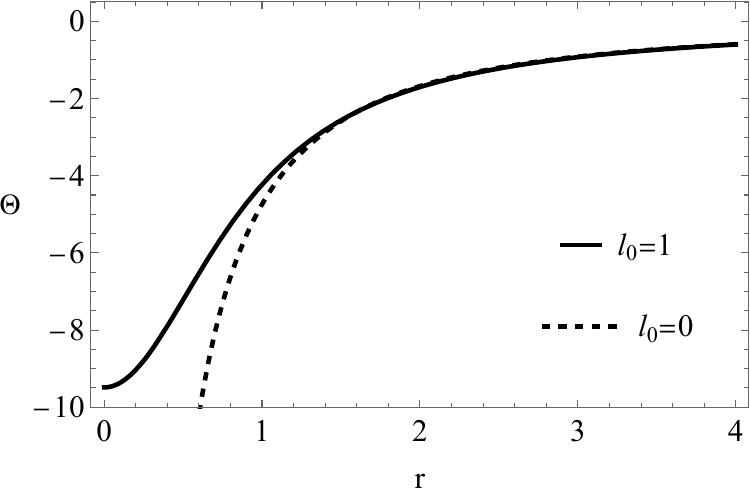}
\caption{Plot of $d \Theta/ d\tau$ and $\Theta$ for $M=5/{\sqrt{G}}$ with $G=1$.	
}
\label{fig:compression}
	\end{figure*}
For the string T-duality metric \eqref{eq:metriccoeff}, we obtain 
\begin{eqnarray}
    \Theta=-\frac{3 \sqrt{2M} (r^2+2 l_0^2)}{2(r^2+ l_0^2)^{7/4}}<0,
    \label{eq:theta}
\end{eqnarray}
which is negative and finite for any $r$. Similarly for the shear tensor the non-zero components are
\begin{eqnarray}
    {\sigma^{r}}_{t}&=& \frac{3 M r^3}{(r^2+l_0^2)^{5/2}}\\
   {\sigma^{r}}_{r}&=&\frac{\sqrt{2 M} r^2}{(r^2+l_0^2)^{7/4}}\\
   {\sigma^{\theta}}_{\theta}&=&{\sigma^{\phi}}_{\phi}=-\frac{\sqrt{2 M} r^2}{2(r^2+l_0^2)^{7/4}}.
\end{eqnarray}
As a result we get
\begin{eqnarray}
    \sigma_{\mu \nu }\sigma^{\mu \nu }=\frac{3 M r^4}{(r^2+l_0^2)^{7/2}} \geq 0,
\end{eqnarray}
which is always finite, positive, and rapidly vanishing ($\sim r^4$) as $r$ goes to zero.
Finally it is straightforward to demonstrate that
\begin{eqnarray}
   R_{\mu \nu }\, u^{\mu} u^{\nu}=\frac{3 M l_0^2 (3 r^2-2 l_0^2)}{(r^2+l_0^2)^{7/2}}.
\end{eqnarray}
We can see that due to the presence of $l_0$ the geometric condition $R_{\mu \nu } u^{\mu} u^{\nu} \geq 0$ can be violated in the region $r < \sqrt{2/3}\, l_0$, but holds in the region $r \geq \sqrt{2/3}\, l_0$. From Einstein's field equations we get
\begin{eqnarray}
    R_{\mu \nu } u^{\mu} u^{\nu}=8 \pi \left( T_{\mu \nu}-\frac{1}{2} T g_{\mu \nu}\right) u^{\mu} u^{\nu} =     \rho+\sum_i P_i \geq 0,\nonumber\\
\end{eqnarray}
where $T_{\mu \nu }u^{\mu} u^{\nu}$ is the (positive definite) energy density of matter as measured by an observer at 4-velocity $u^a$, and $T=g_{\mu \nu }T^{\mu \nu}$ is the trace of the energy-momentum tensor ${T^\mu}_{\nu}=(-\rho, P_r, P_{\theta}, P_{\phi})$.
The above equation is better known as the Strong Energy Condition (SEC).  
Its violation in the region $ r < \sqrt{\frac{2}{3}} l_0 $ results from the presence of stringy effects encoded in the line elements \eqref{eq:metriccoeff}. In principle, the violation of this condition could also occur in the absence of quantum effects, but for the rest of the paper, we will consider only the T-duality case or similar cases where the length scale at which there is a departure from general relativity arises from a quantum gravity formulation.

In the end, we find 
\begin{eqnarray}
   \frac{d \Theta}{d \tau}=-\frac{ 9 M r^2 \left(r^2+(10\,l_0^2)/3 \right)}{2 \left(r^2+l_0^2\right)^{7/2}}\leq 0, \label{dthetatau}
\end{eqnarray}
i.e. a result confirming the convergent character of the geodesics with proper time. However, in contrast to the standard Schwarzschild result (whose compression rate is strictly negative and divergent), in the case above $d\Theta /d \tau$ does not diverge at $r=0$, but goes quadratically to zero. This fact can be seen in Fig. \ref{fig:compression}, where the function $d\Theta/d\tau$ decreases as the radial coordinate decreases, reaching a minimum value at the turning point $r_{\rm tp}\simeq 0.687\,l_0$. 
A repulsive gravity character sets in at smaller $r$, which means that by further decreasing the radial coordinate, the function $d\Theta/d\tau$ increases and approaches zero. 

The turning point $r_{\rm tp}\simeq 0.687\,l_0$ represents the locus of the transition between the outer Schwarzschild geometry and the inner de Sitter geometry, similar to what Sakharov and Gliner obtained with the Israel shell formalism \cite{Gli66,Sak66}. However, in contrast to such attempts, the string T-duality metric has the advantage of a smooth transition without the need to manually impose some junction conditions between the two geometries.

A closer look at \eqref{eq:rayeq} reveals other interesting aspects. 
First, \eqref{eq:rayeq} contains only even functions. This is due to the fact that the transformation $r\to-r$ produces a copy of the spacetime \eqref{eq:metriccoeff} rather than a negative-mass universe.
Second, in the short-scale limit both  $R_{\mu \nu } u^{\mu} u^{\nu}$ and $-\frac{1}{3}\Theta^2$ approach a constant value, while $\sigma_{\mu \nu }\sigma^{\mu \nu }$ vanishes rapidly, as already mentioned.
However, the l.h.s. of \eqref{eq:rayeq} does not become constant in this limit. This can be explained by the fact that the contributions of $R_{\mu \nu } u^{\mu} u^{\nu}$ and $ -\frac{1}{3}\Theta^2$ surprisingly cancel.
Indeed, one finds
\begin{eqnarray}
    - R_{\mu \nu} u^{\mu} u^{\nu}=\Lambda_{\rm eff}-\frac{30 M}{l_0^5}r^2+\dots,
\end{eqnarray}
where $\Lambda_{\rm eff}=6M/l_0^3$ plays the role of an effective cosmological constant.
Similarly 
\begin{eqnarray}
   -\frac{1}{3}\Theta^2=-\Lambda_{\rm eff}+\frac{15 M}{l_0^5} r^2+\dots .
\end{eqnarray}
This behavior is in marked contrast to the Schwarzschild case, which, due to its classical nature, is not able to describe the appearance of a local de Sitter region of quantum origin.

To further examine time like geodesics, we divide \eqref{eq:rayeq} by $\Theta^2$ and integrate over $\tau$.
The net result is
\begin{eqnarray}
    \frac{1}{\Theta(r)}- \frac{1}{\Theta_0}=\frac{\tau}{3}+2\tau \left(\frac{r^4-2 l_0^4 +3 l_0^2 r^2}{3(r^2+ 2 l_0^2)^2} \right),
\end{eqnarray}
where we have fixed the integration constant by using the condition $\Theta(\tau=0)=\Theta_0$.
By using \eqref{eq:theta} we get for the proper time 
\begin{equation}
   \tau=-\frac{(r^2+2 l_0^2)}{r^2}\left(\frac{3(r^2+2 l_0^2)}{\Theta_0 (3 r^2+10 l_0^2)}+\frac{\sqrt{2} (r^2+2 l_0^2)^{7/4}}{ \sqrt{M} (3 r^2+10 l_0^2)}\right).
\end{equation}
As a result, we get a positive proper time for some negative initial compression $\Theta_0<0$.
In such a case, the limit $r\to0$ implies $\tau\to\infty$, i.e., future directed time-like geodesics are complete.
In other words, the existence of an initial trapped surface does not necessarily imply that spacetime will collapse into a curvature singularity. Quantum corrections counteract the full collapse and remove any singular behavior.

Before proceeding, we recall that the present procedure can be extended to a generic cumulative mass distribution profile $ m(r) $ that satisfies regularity conditions at the origin, specifically $ m(r) \sim r^\alpha $ with $ \alpha \geq 3 $ as $ r \to 0 $, and guarantees the existence of a proper definition of the ADM mass at infinity, namely $ m(r) \to \text{constant} $ as $ r \to \infty $. In such a case, we find
\begin{equation}
\label{eq:dthetadtaugeneric}
   \frac{d \Theta}{d \tau}=\frac{m''(r)}{r}-\frac{(m'(r) r)^2-m(r)( 2 m'(r) r-9 m(r))}{2 r^3 m(r)}.
\end{equation}
In this way it is possible to show that the present results for the string T-duality metric are model-independent features that hold for the class of quantum-gravity modified black hole metrics \cite{NSS06,Nic12,IMN13,NiS14,KKM+19}.

It is essential to note that any non-Schwarzschild black hole metric exhibits a two-horizon configuration. In other words, the Schwarzschild geometry represents an unstable configuration in parameter space, where any modifications, such as charge or angular momentum, can trigger the transition to a two-horizon geometry. The ``zero point length'' parameter $ l_0 $ is no exception. However, the local and global behaviors of the manifold are independent concepts, and \eqref{eq:dthetadtaugeneric} captures only the local behavior. This can be easily verified by comparing the Reissner-Nordström geometry with the T-duality metric, both of which exhibit similar global behaviors (two horizons), yet their short-scale behaviors are drastically different, i.e., divergent versus regular.

The case of null geodesics cannot change the conclusion that spacetime is geodesically complete. This can be inferred from the fact that the compression \eqref{eq:theta} and its evolution equation \eqref{dthetatau} depend on the properties of the background spacetime rather than on the nature of the propagating probe, namely massive or massless.
Nevertheless, we briefly sketch the reasoning here.

We start by considering a simplified picture, namely the two regions mentioned above, an inner de Sitter core for $r\simeq l_0$ and an outer Schwarzschild geometry for $r\gg l_0$. We also assume that the transitions between the above two occur on a spacetime layer, but its details are unimportant for the present discussion.
In short, the problem is reduced to two known problems, namely the determination of the compression in the above geometries. Ultimately, geodesic completeness is related to what happens near the origin, so what happens at large distances is irrelevant, and the problem reduces to studying the de Sitter core. 

To this end, we recall that a spacetime of constant curvature deforms wave propagation much like a mass term. 
For example, the photon field wave operator is modified as 
\begin{equation}
\Box\longrightarrow \nabla_\alpha\nabla^\alpha-R
\end{equation}
where $\Box$ is the flat space d'Alembert operator and $R=R_{\mu\nu}\ \delta_{\mu\nu}$, with $R_{\mu\nu}\sim \Lambda_\mathrm{eff}\ \delta_{\mu\nu}$ is the Ricci tensor. This means that for photons the study of null geodesics becomes equivalent to that of time-like geodesics.

For the scalar field, the situation is similar but not identical. Since the scalar fields can be minimally coupled to gravity, we have to rely on general results for null geodesics in de Sitter space.
In this context, we recall that null curves in de Sitter space have been the subject of a large literature -- see for example \cite{GaS07,LoO09,LoO12,Cot17}. Despite the presence of null lines and points that cannot be connected, de Sitter space is said to be geodesically complete \cite[p. 126]{HaE73} (see also \cite{KOCP02}).

\section{Conclusions}
\label{sec:conclusions}

The scrutiny of the Raychaudhuri equation reveals a surprising new nature of gravity in the presence of T-duality effects. While the convergent character of geodesics remains unchanged ($\Theta$ stays negative), the degree of convergence $d\Theta/d\tau$ is decisively affected by quantum gravity effects. At short scales, gravity changes its character!  Quantum gravity is actually repulsive and weakens the geodesic compression, while the conventional attractive character is just a large distance effect of the classical gravity regime. This attractive character is spurious because it is not connected to the fundamental quantum nature of gravity, but rather results similarly to  the Debye screening in electrostatics. 
Instead of charge carriers, the quantum vacuum acts like a polarizable medium in the case of gravity and is responsible for the screening. At large distances, the quantum vacuum ``collapses'' into the classical vacuum and the attractive regime sets in. %
In case of electric forces, the bare charge is actually larger (in magnitude) than the dressed one; for gravity, however, the bare quantum force is weaker, and also has a different sign!


Another important point is that the character of quantum gravity cannot depend on the global properties of spacetime, such as symmetries and other features of the classical gravity regime. In fact, the attractive nature of quantum gravity depends on the length scale $ l_0 $, i.e., on local spacetime properties. This confirms that the genuine characteristic of quantum gravity is the intrinsic loss of resolution at the Planck scale, which remains unaffected regardless of the overall structure of the metric at large distances. 
An illustrative example can also be found in the regime of semiclassical gravity, where the conformal anomaly of quantum fields is a purely geometric local object arising from renormalization in curved space. It does not depend on quantum states over the manifold, which instead rely on global properties, such as the existence of a suitable coordinate covering \cite[p. 178]{BiD84}.

Turning to the specifics of our analysis,  \eqref{eq:dthetadtaugeneric} does not rely on global properties, as it is sensitive to the short-distance behavior of the manifold. This sensitivity is highlighted by the scalar nature of the compression parameter $ \Theta $. The reasoning reinforces the idea that our conclusions can be extended to other spacetimes without spherical symmetry, which means that our results also hold for other spacetime configurations, such as the state of the universe during the Planck epoch.

Finally, we emphasize that the scenario outlined above is general and model-independent. While alternative formulations of   
quantum gravity may predict slightly different values for some parameters, they cannot alter our conclusions: classical gravity  
gravitates, but quantum gravity antigravitates.


\begin{acknowledgments}
The work of P.N. has partially been supported by GNFM, Italy's National Group
for Mathematical Physics.
\end{acknowledgments}

\end{document}